\documentclass{article}

\usepackage{arxiv}
\usepackage{enumitem}
\usepackage[utf8]{inputenc} 
\usepackage[T1]{fontenc}    
\usepackage{url}            
\usepackage{hyperref}       
\usepackage{booktabs}       
\usepackage{amsfonts}       
\usepackage{nicefrac}       
\usepackage{microtype}      
\usepackage{color}
\usepackage{xcolor}
\usepackage{lipsum}
\usepackage{graphicx}
\usepackage{amsmath}
\usepackage{booktabs}
\usepackage{multirow}
\usepackage{array}
\usepackage{arydshln} 
\usepackage{makecell}
\usepackage{soul}
\graphicspath{ {./images/} }
\usepackage{wrapfig}  
\usepackage{subcaption}
\usepackage[style=numeric,sorting=none,maxbibnames=5]{biblatex}
\DeclareFieldFormat[misc]{title}{"#1"}
\title{Addressing the sustainable AI trilemma: a case study on LLM agents and RAG}

\author{
 Hui Wu \\
  Dept. of Engineering\\
  University of Exeter\\
  Exeter, UK  \\
  EX4 4QF \\
  \texttt{hw865@exeter.ac.uk} \\
   \And
  Xiaoyang Wang \\
  Dept. of Computer Science\\
  University of Exeter\\
  Exeter, UK  \\
  EX4 4QF \\
  \texttt{x.wang7@exeter.ac.uk} \\
  \And
  Zhong Fan \\
  Dept. of Engineering\\
  University of Exeter\\
  Exeter, UK \\
  EX4 4QF \\
  \texttt{z.fan@exeter.ac.uk} \\
}

\begin{document}
\maketitle
\begin{abstract}

Large language models (LLMs) have demonstrated significant capabilities, but their widespread deployment and more advanced applications raise critical sustainability challenges, particularly in inference energy consumption. We propose the concept of the Sustainable AI Trilemma, highlighting the tensions between AI capability, digital equity, and environmental sustainability. Through a systematic case study of LLM agents and retrieval-augmented generation (RAG), we analyze the energy costs embedded in memory module designs and introduce novel metrics to quantify the trade-offs between energy consumption and system performance. Our experimental results reveal significant energy inefficiencies in current memory-augmented frameworks and demonstrate that resource-constrained environments face disproportionate efficiency penalties. Our findings challenge the prevailing LLM-centric paradigm in agent design and provide practical insights for developing more sustainable AI systems. Our code is available online\footnote{\url{https://github.com/huiwxing/llmagent_trilemma}}. 

\end{abstract}


\section{Introduction} 
Large language models (LLMs) have played an important role in artificial intelligence, significantly advancing natural language understanding and generative AI. These models are now powering applications across healthcare, education, and business, demonstrating their growing practical impact in everyday services. However, this rapid development comes with significant challenges, particularly in terms of energy use and its impact on the environment and society. While the energy consumption of LLM training has received much attention \cite{patterson2022carbon} , recent evidence \cite{chien2023reducing} \cite{desislavov2023trends} \cite{jiang2024preventing} suggests that the inference phase, particularly in large-scale deployments, has emerged as the dominant contributor to the overall energy footprint, highlighting a critical and growing concern in the sustainability of LLM-powered services.

In response to these challenges, we propose the concept of the \textbf{Sustainable AI Trilemma}, which draws parallels to the sustainability trilemma balancing Economy, Society, and Environment \cite{wu2024triple}. In the AI context, this trilemma highlights the trade-offs among \textbf{capability, digital equity, and environmental sustainability}. Current LLM deployments heavily prioritize capability—pushing for greater model performance and broader applications—often at the expense of equity and sustainability. This imbalance is made worse by the increasing use of \textbf{LLM-based agents}, and the scaling of \textbf{multi-agents}, which applies LLMs in a frequent manner for long-term operations.

LLM-based agents, capable of executing complex, multi-step tasks, are increasingly integrated across domains such as enterprise automation, scientific research, and personalized assistance \cite{xi2023risepotentiallargelanguage}. This paradigm amplifies the energy demands of LLM inference, raising urgent questions about the sustainable scalability of such AI services. Without deliberate attention to energy efficiency, the widespread use of these agents risks accelerating resource inequities \cite{khowaja2024chatgpt} and undermining global sustainability goals \cite{faiz2023llmcarbon}.

Despite these challenges, most research on LLM agents focuses on improving performance, with little consideration of sustainability or equity metrics. As a first step toward addressing the trilemma, we use LLM agents as a case study to better understand these challenges. This paper aims to address this gap with the following contributions:

\textbf{Proposing the Sustainable AI Trilemma and Highlighting its Urgency:}
We articulate the risks posed by an overemphasis on productivity in LLM agent deployments, which exacerbates the trilemma by neglecting equity and environmental sustainability.

\textbf{Unified Analysis of Energy Cost in LLM Agents:} Through systematic analysis and a case study of the memory module, we reveal the energy costs embedded in LLM agent and RAG designs. Our findings emphasize the trade-offs between enhanced functionality and sustainability, demonstrating that resource-intensive designs often yield diminishing performance returns and identifying opportunities for more sustainable implementations.

\textbf{New Metrics for Energy and Performance Trade-offs:} We introduce novel metrics to evaluate the trade-offs between energy cost and performance in LLM agents. These metrics provide a foundation for optimizing the design and operation of LLM agents to achieve more energy efficiency without compromising effectiveness.

\textbf{Challenging the Paradigm of LLM-Centric Autonomy:}
A growing trend in agent design involves offloading all subtasks to the LLMs, purportedly to enhance autonomy. From the perspectives of environmental impact and cost-efficiency, we challenge this prevailing trend, demonstrating through empirical evidence that this approach often results in unnecessary computational overhead without proportional benefits in agent autonomy or performance, relying excessively on LLMs undermines the broader objectives of sustainable AI.

The remainder of this paper is organized as follows: Section 2 introduces the Sustainable AI Trilemma and its four constituent dilemmas, along with background on LLM agents and LLM energy consumption. Section 3 presents a case study on memory modules in LLM-based agents, examining their impact on energy efficiency and system complexity, and proposes metrics for evaluating energy-performance trade-offs. Section 4 presents our experimental results and analysis. Finally, Section 5 discusses implications and concludes with insights for future sustainable AI development.

\section{Background}
\label{sec:headings}

\subsection{Sustainable AI Trilemma}
The concept of a trilemma, which highlights the difficulty of balancing competing priorities, has been explored in various contexts, including sociology and ethics. Recent sociological studies have applied this trilemma framework to AI \cite{ernst2022ai}, and others have emphasized the sustainability of AI \cite{falk2024challenging}\cite{10102375}. The AI trilemma posits that AI development must balance three fundamental goals: AI capability or performance, sustainability, and equity, where optimizing for any two often comes at the expense of the third. 
However, with the rapid rise of LLMs and their pervasive influence, we argue that this trilemma must be redefined and addressed from a technical perspective. The Sustainable AI Trilemma emerges from a series of intertwined technical dilemmas that reflect conflicting priorities in the design, deployment, and long-term operation of AI systems. Below, we identify and analyze four core dilemmas, each highlighting a critical technical challenge.
\begin{figure}[htbp]
    \centering
    \includegraphics[width=0.75\textwidth]{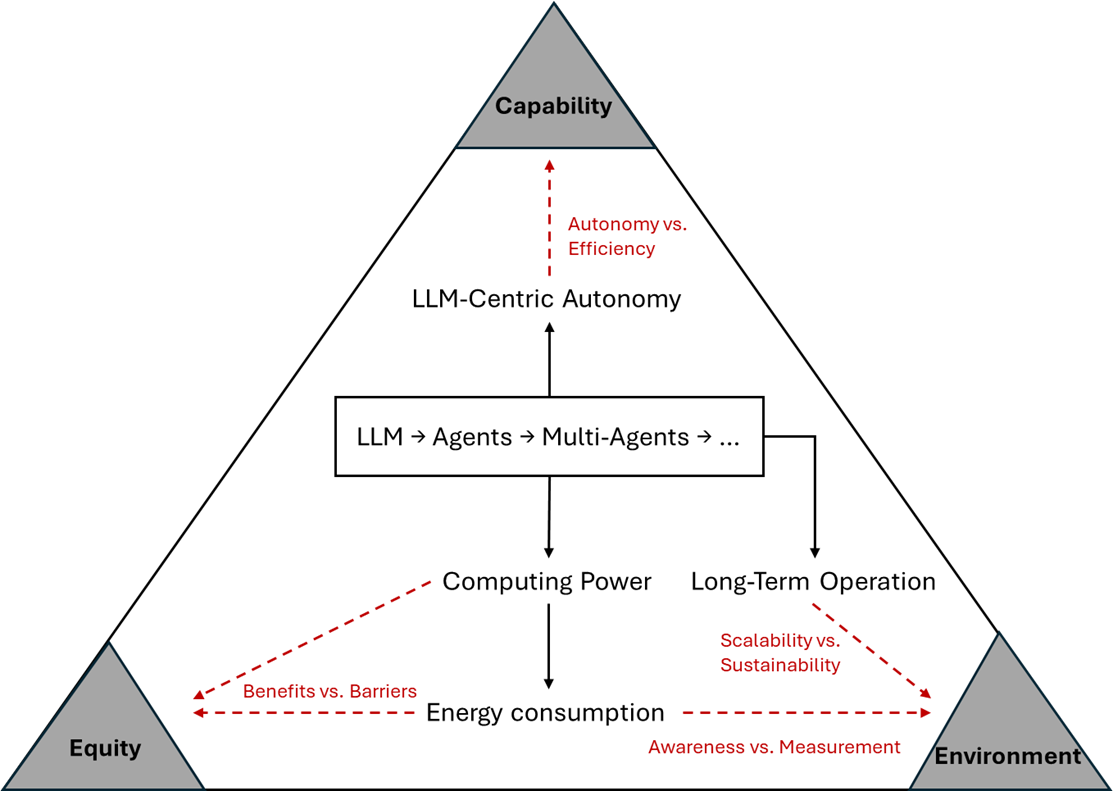}
    \caption{The Sustainable AI Trilemma.}
\end{figure}
\paragraph{Evaluation Dilemma: Awareness vs. Measurement}
While the environmental impacts of LLMs and other AI systems have become a recognized concern, the lack of accurate and comprehensive assessment tools hinders progress toward sustainability. Existing tools, such as MLCO2~\cite{lacoste2019quantifying} and Green Algorithms~\cite{GreenAlgorithms}, are designed for traditional machine learning and earlier AI algorithms. These tools fail to account for the unique challenges posed by LLMs and their applications, such as memory-intensive inference processes, highly variable usage patterns, and the lifecycle emissions from large-scale deployment. This technical gap leaves practitioners and policymakers uncertain about the true scale of AI’s environmental footprint, creating a paradox where actionable insights are hindered despite widespread recognition of the problem.

\paragraph{Design Dilemma: Autonomy vs. Efficiency}
The pursuit of highly autonomous AI systems \cite{xi2023risepotentiallargelanguage}, exemplified by the rise of single- and multi-agent frameworks, drives the demand for increasingly sophisticated architectures. While these systems promise greater flexibility and functionality, their designs often introduce redundancy and inefficiencies \cite{LIU2025112278}, such as poorly optimized workflows leading to excessive and unnecessary model calls, frequent over-reliance on LLMs for subtasks that could be addressed more efficiently with specialized modules, or maintaining multiple concurrent LLM instances for similar tasks across different services. This tradeoff between expanding autonomy and maintaining energy-efficient operations adds strain to system-level resource management, resulting in diminishing returns in capability.
\paragraph{Operational Dilemma: Scalability vs. Sustainability}
Scaling AI services to meet increasing demand exacerbates the tension between scalability and sustainability. Expanding the number of deployed LLM-based agents requires maintaining consistent Quality of Service (QoS) under growing workloads. However, the infrastructure required to ensure QoS—such as low-latency inference, real-time responsiveness, aggressive caching and pre-warming strategies \cite{ghosh2024enablingefficientserverlessinference}—inevitably leads to heightened energy consumption and operational costs. At scale, these architectural inefficiencies can be further amplified. While traditional AI systems benefit from MLOps and AIOps \cite{10.1145/3625289} to streamline operations, these tools face significant challenges when applied to the complexity of LLM applications. The memory-intensive processes, cross-agent interactions, and continuous learning demands of LLM-based systems strain the existing operational workflows, raising questions about their scalability and long-term feasibility in resource-constrained environments.
\paragraph{Access Dilemma: Benefits vs. Barriers}
Resource-intensive AI applications disproportionately benefit well-resourced organizations and regions while marginalizing underprivileged communities. Proprietary LLMs often rely on expensive API-based access models, imposing high financial costs on users. Open-source LLMs, while enabling local deployment, demand highly efficient hardware to mitigate their energy-intensive operations, which are often unavailable in resource-limited regions. These twin pressures—economic costs and environmental impacts—place an outsized burden on communities with limited access to affordable and reliable computing infrastructure. The resulting disparities exacerbate the digital divide \cite{khowaja2024chatgpt}, effectively excluding these communities from accessing and benefiting from advanced AI services that could otherwise drive meaningful societal and economic change \cite{yu2024chatgptunveilingrealworldeducational}\cite{baguma2023examining}.

These four dilemmas collectively form the foundation of the \textbf{Sustainable AI Trilemma}: the need to balance AI capability, digital equity, and environment. To address this trilemma, we believe a paradigm shift in designing, deploying, and evaluating AI systems is required.

\subsection{LLM-based Agents}
The emergence of LLM-based agents represents a transformative shift in the application of large language models. Unlike traditional applications that rely on single-turn interactions, LLM-based agents are designed to perform multi-turn dialogues and execute complex, multi-step tasks autonomously \cite{cheng2024exploring}. These agents are increasingly deployed across diverse sectors, from intelligent customer support systems to advanced research assistants, showcasing their potential to revolutionize workflows and decision-making processes. At their core, LLM-based agents integrate multiple functional components to achieve greater autonomy and adaptability. Key components typically include:

\begin{itemize}[left=0pt]
\item \textbf{Planning Module}: Breaks down high-level tasks into manageable sub-tasks, enabling structured and goal-oriented operations \cite{huang2024understanding}.
\item \textbf{Tool Use}: Incorporates external APIs, databases, or specialized software tools to enhance problem-solving capabilities \cite{shen2024llm}.
\item \textbf{Memory Module}: Allows agents to retain and retrieve past information, supporting contextual understanding and long-term consistency.
\end{itemize}

Among these, the memory module is particularly important. While traditional LLM applications depend solely on the pretrained knowledge embedded within the model and the immediate context provided by user queries, memory-enabled LLM agents or Retrieval-Augmented Generation (RAG) \cite{zhao2024retrieval} extend this paradigm by incorporating both internal memory (e.g., conversation histories) and external memory (e.g., indexed document repositories) \cite{zhang2024surveymemorymechanismlarge}. This memory capability facilitates advanced applications by mitigating LLM hallucination \cite{zhang2023siren}, providing improved performance through enriched context and persistent knowledge access.

\subsection{Energy Cost and Environmental Impacts of LLM Inference}
The growing energy consumption of AI systems, particularly LLMs, has become a significant concern in the field and inference-phase energy consumption deserves particular attention, especially in terms of carbon footprints \cite{10765824}. \cite{desislavov2023trends} suggests that the multiplicative effect due to widespread AI deployment remains a crucial consideration. Studies in \cite{chien2023reducing} have shown that for services like ChatGPT, inference-related carbon emissions in one year amount to 25 times that of GPT-3's training. This trend has driven the development of specialized measurement tools and analysis. Green Algorithms \cite{GreenAlgorithms} provides an online platform for estimating computational carbon footprints, while LLMCarbon \cite{faiz2023llmcarbon} specifically addresses LLM scenarios, offering end-to-end carbon emission predictions for both dense and mixture-of-experts (MoE) architectures.

Empirical studies have extensively investigated LLM inference energy consumption. \cite{WordsWatts} have conducted large-scale testing of LLaMA models across different GPU platforms (V100 and A100), including scenarios with up to 32-GPU model sharding. Through comprehensive characterization studies of mainstream LLMs, Wilkins et al. \cite{wilkins2024offline} have developed high-accuracy ($R^2>0.96$) models for energy consumption and runtime prediction. Further research \cite{argerich2024measuring} has explored how factors such as model size, layer count, parallel attention mechanisms affect energy consumption, along with optimization techniques like batch processing and quantization. In code generation, Vartziotis et al. \cite{vartziotis2024learn} have evaluated the sustainability awareness of commercial AI models like GitHub Copilot, ChatGPT-3, and Amazon CodeWhisperer. What's more, the optimization of AI energy consumption requires careful consideration of trade-offs - while some AI models aim to reduce carbon emissions, their training and deployment also generate environmental impacts, yielding positive net benefits only under appropriate scenarios \cite{DELANOE2023117261}.

\section{Computational and Energy Costs in LLM-based Agents: A Case Study on Memory Modules}
The memory module is a transformative feature in LLM agents. However, the incorporation of memory introduces notable challenges in energy efficiency and design complexity. In this section, we take the memory module of LLM-based agents as an example to explore potential inefficiencies in resource utilization within their design and operations.
\subsection{Problem Definition}
\label{sec: problem}
To understand its impact, we first define the simplest case of an LLM agent without memory before formalizing the enhanced capabilities of a memory-augmented agent.
\paragraph{Without Memory}
In its simplest configuration, an LLM agent operates as a standalone inference system. Given a user query $Q$, the LLM generates a response $A$:
\begin{equation}
LLM(Q)=A
\end{equation}
This formulation assumes no access to external data or stored context. The model relies entirely on its pretrained knowledge and immediate input $Q$. While computationally efficient, this approach is limited by the LLM's inherent knowledge boundaries, which cannot be updated without retraining.
When $Q$ references information outside the model's scope, the output $A$ is prone to inaccuracies or hallucination, failing to meet the requirements of knowledge-intensive or context-dependent applications.

\paragraph{With Memory}
When a memory module is incorporated, the memory $M$ is typically categorized into internal memory and external memory, i.e., $M = \{ M_{internal}, M_{external}\}$. Internal memory comprises historical information collected during the agent’s operation. For example, an agent solving iterative tasks may store user inputs $i$ and its own responses $o$ from prior interactions at time $t$: $M_{internal} = \{(i_1, o_1), (i_2, o_2), \dots, (i_{t-1}, o_{t-1})\}$. External memory refers to data retrieved from domain-specific repositories, databases, documents, or other external sources: $M_{external} = \{D_1, D_2, \dots, D_n\}$, where these external resources $D$ are indexed and queried as needed.
The behavior of the memory-enabled agent can be described as:
\begin{equation}
LLM(Q,M_q)=A
\end{equation}
Here, $M_q\in M$ represents the relevant subset of the memory $M$, retrieved based on the query $Q$. We assume that for every query there is a ground truth answer $A_G$, which can be generated based on their ground truth memories $M_G$.

\paragraph{Energy Model} Prior work \cite{wilkins2024offline} has established an analytical model for LLM inference energy consumption that accounts for both input and output text lengths, as well as their interaction. Specifically, the energy consumption $E$ of a primary input-output of a certain LLM $K$ can be expressed as
\begin{equation}
 E= e_{K} (T_{in}, T_{out}) = \alpha_{K,0} *T_{in} + \alpha_{K,1}*T_{out} + \alpha_{K,2}*T_{in}*T_{out}
\end{equation}
where $T_{in}$ and $T_{out}$ represent the number of input and output tokens respectively, and $\alpha_{K,0}$, $\alpha_{K,1}$, and $\alpha_{K,2}$ are model-specific ($K$) coefficients. Since output token generation is more energy-intensive than input processing \cite{wilkins2024offline} , $\alpha_{K,1}$ normally exceeds both $\alpha_{K,0}$ and $\alpha_{K,2}$. We will use this model to perform some theoretical simulation analyses in the next section and some metrics calculations in Experiment ~\ref{sec:Task 2}.

\subsection{Cost of Existing Memory Module Design Patterns}

\begin{figure}[htbp]
    \centering
    \includegraphics[width=0.8\textwidth]{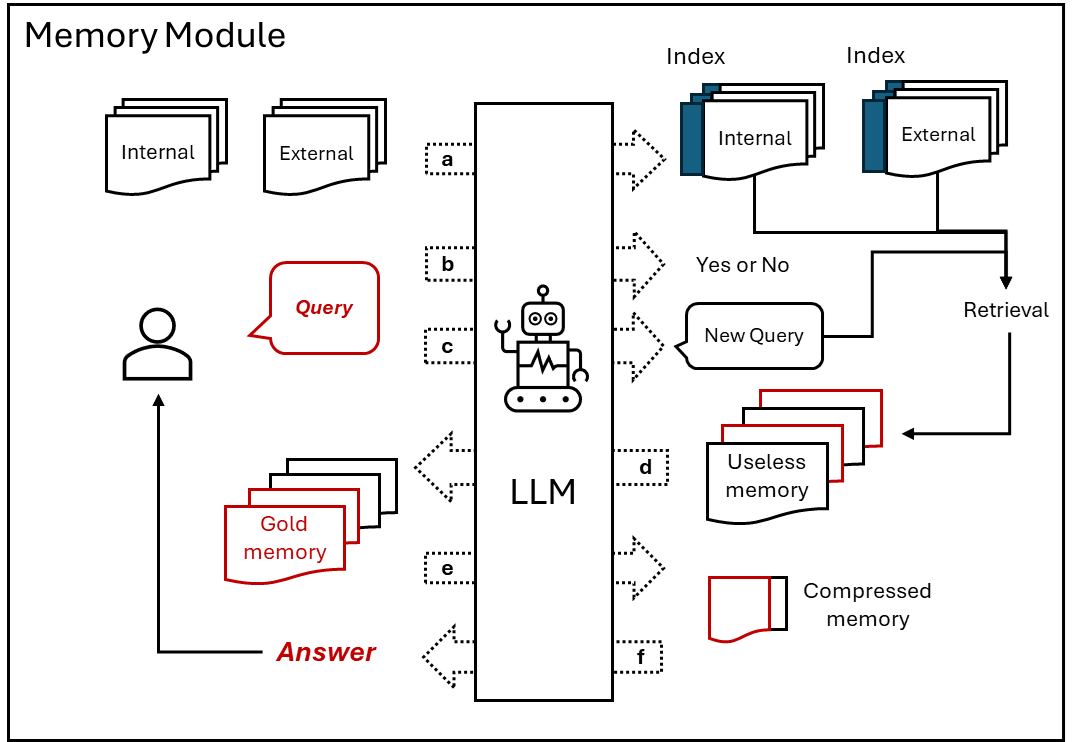}
    \caption{Design of Memory Modules in LLM Agents or RAG. The figure contains the operations that require LLM inference: (a) Memory Formation, (b) Retrieval Necessity Detection, (c) Query Optimization, (d) Reranking, (e) Compression, (f) Generation. The red markers represent data that is necessary for a QA process to generate a correct response, while black indicates non-necessary data.}
    \label{fig:memory module}
\end{figure}

The integration of memory modules introduces more challenges. Many existing memory designs completely rely on LLM operations (as shown in Figure ~\ref{fig:memory module}), prioritizing performance and autonomy improvements without considering computational workloads and resource consumption. Through analyzing different components of memory-augmented frameworks, we identify three critical phases in modern LLM memory architecture: (a) \textbf{Memory Formation}, which processes and organizes internal/external input data into storage structures; (b-c) \textbf{Memory Reading}, encompassing retrieval necessity detection and query optimization for better retrieval; and (d-f) \textbf{Memory utilization}, to generate better answers by optimizing memory usage. This paper aims to uncover the hidden costs of these components and quantify the trade-offs between enhanced task performance and increased energy consumption.

\subsubsection{Memory Formation}
Memory formation involves the transformation of raw data into structured and retrievable representations using various indexing and storage optimization techniques. Vector-based storage and retrieval, which transforms raw data into vector embeddings for efficient similarity-based searches, struggles with scalability and complex query requirements such as comprehensive full-text searches, relation-based graph, and keyword search \cite{jing2024large}. Several alternative LLM indexing strategies have been proposed. Keyword or topic-based indexing, implemented in systems like MemoChat \cite{lu2023memochat} and MemoryBank \cite{zhong2024memorybank}, leverages LLMs to extract salient terms from the input data. Summary-based indexing, adopted by systems such as SCM \cite{wang2023enhancing}, MemGPT \cite{packer2023memgpt}, and MPC \cite{lee-etal-2023-prompted}, employs LLMs to generate concise representations of memory segments. A more sophisticated approach is to build structured representations, such as triplets and knowledge graphs, by extracting entities and relationships through LLM. Systems like Think-in-Memory \cite{liu2023think} and RET-LLM \cite{modarressi2024retllm} exemplify this approach. 

While advanced memory representations excel in supporting complex reasoning tasks, they impose significant computational overhead due to multiple LLM processing iterations, raising concerns about scalability and efficiency as memory data volumes grow. This raises important questions about the trade-off between costs and performance benefits. For instance, when most of indexed memories see limited or no retrieval, the substantial energy investment in memory formation processes becomes difficult to justify against the minimal practical benefits. Such scenarios highlight the critical challenge of aligning indexing strategies with specific application requirements to ensure optimal resource utilization.

\subsubsection{Memory Reading}
\label{sec:Memory Reading}
Through the preprocessing process of Memory Formation, the LLM agent is now equipped with a retrievable memory/knowledge base. Memory Reading, on the other hand, is the process of receiving a query and then getting a portion of information from the base that matches the query.

\paragraph{Retrieval Necessity Detection}
While RAG improves response accuracy and reduces hallucinations, persistent indiscriminate retrieval can cause unnecessary overhead. Retrieval necessity detection optimizes efficiency by using the LLM's inherent knowledge for simple tasks and enabling selective memory access when necessary. SCM \cite{wang2023enhancing} and Ren et al. \cite{ren2023investigating} design prompt like Self-Asking prompt and Judgment prompt to assess memory activation needs. MemGPT \cite{packer2023memgpt} treats LLM as an operating system and empowers the LLM processor to autonomously decide, update, and search its memory based on the current context. Rowen \cite{ding2024retrieve} introduces a multilingual semantic-aware detection module that evaluates cross-language response consistency and triggers retrieval augmentation when factual inconsistencies are detected. SKR \cite{wang-etal-2023-self-knowledge} uses both direct prompting and training a smaller classifier, to detect whether the model has ”self-knowledge”.

Building on these studies, we distill the Retrieval Necessity Detection process to its fundamental form, as outlined below in terms of the LLM inference operations (To simplify the analysis, system prompt templates in the input are temporarily excluded. The following analysis remains consistent.):

\begin{itemize}[left=0pt]
\item \textbf{Input}: Query (token number = $T_q$).
\item \textbf{Output}: A boolean or numeric score (indicating whether retrieval is needed, token number = $C$, where $C$ is a small constant).
\item \textbf{Energy}: $E_1= e_K(T_q, C) = \alpha_{K,0} *T_q + \alpha_{K,1}*C + \alpha_{K,2}*T_q*C$
\end{itemize}

\paragraph{Query Optimization}
Query optimization enhances retrieval accuracy by enriching raw or under-contextualized queries with semantic information through query expansion or transformation. Query expansion generates additional queries derived from the original input like decomposing complex questions into sub-questions, or employing prompt engineering techniques to create multiple contextually enriched variations, such as the Least-to-most prompting \cite{zhou2023leasttomostpromptingenablescomplex}, Chain-of-Thought \cite{wei2023chainofthoughtpromptingelicitsreasoning}, Self-Ask \cite{press-etal-2023-measuring}, and Multi Query Retrieval \cite{langchain2024multiquery}. Query transformation focuses on rewriting or reinterpreting the original query to better align with retrieval objectives, such as Hypothetical Document Embedding (HyDE) \cite{gao-etal-2023-precise}, Step-back prompting \cite{zheng2024take}, Rewrite-Retrieve-Read \cite{ma-etal-2023-query}. Both approaches increase the semantic richness and correctness of the query, enabling retrieval systems to access a broader and more relevant information space. However, it is important to note that these enhancement methods typically not only increase the token count of the original query but also represent an additional computational cost in themselves.

The basic form of query optimization is as follows:
\begin{itemize}[left=0pt]
\item \textbf{Input}: Query (token number = $T_q$).
\item \textbf{Output}: New query/queries, or hypothetical documents (token number = $T_{qn}$, which is usually larger than $T_q$).
\item \textbf{Energy}: $E_2= e_K(T_q, T_{qn}) = \alpha_{K,0} *T_q + \alpha_{K,1}*T_{qn} + \alpha_{K,2}*T_q*T_{qn}$
\end{itemize}

\paragraph{Memory Retrieval}
Memory retrieval is the core step where relevant memory entries are extracts based on the transformed query, typically using similarity metrics (e.g., cosine similarity) between query and document embeddings. Methods include sparse retrieval (e.g., BM25 \cite{robertson2009probabilistic}), dense retrieval (e.g., Sentence-BERT \cite{reimers2019sentencebertsentenceembeddingsusing}), or hybrid retrieval combining both. Although some approaches like Memochat \cite{lu2023memochat} attempt to use LLMs as retrievers, we argue that such methods are significantly limited by the LLM's context window size and lack scalability in retrieval speed and volume, making them unsuitable for consideration in this study. As a fixed computational cost \textbf{independent of LLM inference}, memory retrieval is considered an essential step in the pipeline.

\subsubsection{Memory Utilization}
\label{sec:Memory Utilization}
Memory utilization involves additional processing of retrieved memory, such as re-ranking and compression, to integrate the memory with the query in an appropriate format as input to LLM for generating answers.

\paragraph{Reranking}
Reranking serves as a refinement step in the RAG process, where retrieved documents are reordered based on their relevance to the query. Recent advancements have increasingly focused on LLM-based reranking. Zhuang et al. \cite{zhuang2023opensourcelargelanguagemodels} demonstrated that LLMs pre-trained on unstructured text, like LLaMA and Falcon, exhibit strong zero-shot ranking capabilities. These methods such as LRL \cite{ma2023zeroshotlistwisedocumentreranking}, UPR \cite{sachan-etal-2022-improving}, RankGPT \cite{sun-etal-2023-chatgpt}, PRP \cite{qin-etal-2024-large} leverage prompt engineering with models like GPT-3/4, UL2, and T5 for zero-shot passage reranking, employing various prompt types, including query generation, relevance genenration \cite{liang2023holisticevaluationlanguagemodels}, pairwise ranking, and permutation for ranked lists. Although prompt-based approaches reduce fine-tuning overhead and enhance agent autonomy, they themselves introduce extra computational costs during inference by processing multiple retrieved documents simultaneously. However, similar to Retrieval Necessity Detection, considering that reranker outputs are typically concise (ranked indices or scores, a small constant of output token numbers), these approaches have the potential to be cost-effective with moderate number of documents retrieved.

The basic form of LLM reranker is as follows (if using permutation generation prompting):

\begin{itemize}[left=0pt]
\item \textbf{Input}: New query or queries, Retrieved related memories (token number = $T_{qn}+T_m$).
\item \textbf{Output}: Ranked index (token number = $K$, where $K$ represents the number of top-K retrieved entries).
\item \textbf{Energy}: $E_3= e_K(T_{qn}+T_m, K) = \alpha_{K,0} *(T_{qn}+T_m) + \alpha_{K,1}*K + \alpha_{K,2}*(T_{qn}+T_m)*K$
\end{itemize}

\paragraph{Compression}
After retrieval and reranking, relevant documents may still exceed LLM input capacity and introduce noise. While compression strategies can help by extracting essential information, current approaches face significant trade-offs. LLM-based summarization and specialized compression models (e.g. PRCA \cite{yang2023prcafittingblackboxlarge}, RECOMP \cite{xu2023recompimprovingretrievalaugmentedlms}) achieve good performance, but generate significantly more output tokens and consume more energy (owing to the large output coefficient $\alpha_{K,1}$ and the number of output tokens $\beta*T_m$, and the square interaction term $\beta*T_m^2$ ) than simpler steps like retrieval detection or reranking. Alternative entropy-based methods like Selective Context \cite{li2023compressingcontextenhanceinference} and LLMLingua \cite{jiang2023llmlinguacompressingpromptsaccelerated} reduce costs by avoiding full LLM inference, but may compromise compression quality \cite{pan2024llmlingua2datadistillationefficient}. The overall efficiency gains versus computational overhead remains a critical consideration, especially at scale.

The basic form of LLM compressor is as follows:
\begin{itemize}[left=0pt]
\item \textbf{Input}: Ranked memories, which has the same token as unranked just in a different order (token number = $T_m$).
\item \textbf{Output}: Compressed memories (token number = $\beta*T_m$, $0<\beta<1$, where $\beta$ is the compression ratio).
\item \textbf{Energy}: $E_4= e_K(T_m, \beta*T_m) = \alpha_{K,0} *T_m + \alpha_{K,1}*\beta*T_m + \alpha_{K,2}*\beta*T_m^2$
\end{itemize}

\paragraph{Generation}
The final step synthesizes the query and compressed memory to produce a coherent answer. The input query and output answer are unavoidable computational costs, making compressed memories the key variable factor. While a larger compression ratio ($\beta$) increases energy costs, it also preserves more information, potentially leading to more accurate and complete answers. The optimal compression ratio thus needs to balance the trade-off between computational efficiency and answer quality based on specific application requirements.

\begin{itemize}[left=0pt]
\item \textbf{Input}: Original query (token number = $T_q$), Compressed memories (token number = $\beta*T_m$, $0<\beta<1$) .
\item \textbf{Output}: Answer (token number = $T_a$).
\item \textbf{Energy}: $E_5= e_K(T_q+\beta*T_m, T_a) = \alpha_{K,0} *(T_q+\beta*T_m)+ \alpha_{K,1}*T_a + \alpha_{K,2}*(T_q+\beta*T_m)*T_a$
\end{itemize}

\subsubsection{Memory Management}
Beyond basic memory read and write operations, some research focuses on the long-term maintenance and management of memory in LLM agents. These approaches often emphasize agent autonomy, delegating much of the memory management process to the LLM itself. For instance, Think-in-Memory \cite{liu2023think} introduces prompts designed specifically for memory merging and forgetting. These methods also introduce more uncertainties in energy consumption over long-term deployment. Further exploration in our future research to balance autonomy and sustainability in memory management.

\subsection{Metrics for Energy and Performance Trade-offs}
In the evaluation of RAG systems, assessment tasks are typically divided into two phases: \textbf{Retrieval} and \textbf{Generation}. Corresponding to the evaluation subjects defined in Section~\ref{sec: problem}, we consider the following elements: query ($Q$), answer ($A$), ground truth answer ($A_G$), retrieved relevant memories ($M_q$), and ground truth memories ($M_G$). Yu et al. \cite{yu2024evaluation} synthesized current research on RAG evaluation and categorized various metrics into five fundamental types: $\textbf{Relevance}(M_q \leftrightarrow Q)$ and $\textbf{Accuracy}(M_q \leftrightarrow M_G)$ in 
the Retrieval phase, $\textbf{Relevance}(A \leftrightarrow Q)$, $\textbf{Faithfulness}(A \leftrightarrow M_q)$, and $\textbf{Correctness}(A \leftrightarrow A_G)$ in the Generation phase, where $\leftrightarrow$ denotes the alignment or relationship between two elements (e.g., $M_q \leftrightarrow Q$ represents how relevant the retrieved documents are to the query).

\begin{figure}[htbp]
    \centering
    \includegraphics[width=0.8\textwidth]{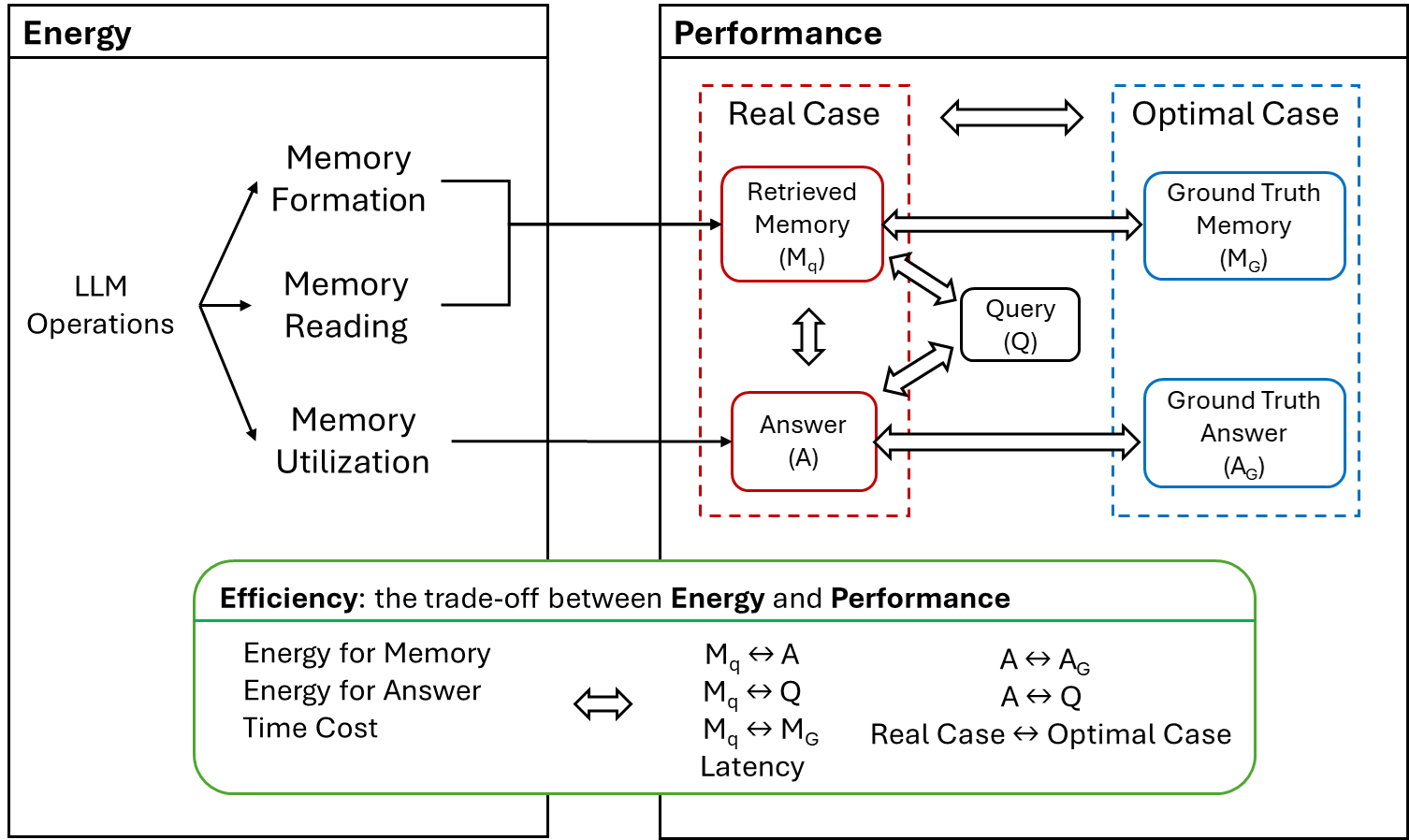}
    \caption{New efficiency metrics to balance energy cost and task performance.}
\end{figure}
\subsubsection{Trade-offs Between Memory Formation and Retrieval}
The choice of indexing methods during memory formation not only affects system energy consumption but also directly impacts subsequent retrieval efficiency and effectiveness. To comprehensively evaluate different indexing strategies, we propose a multi-dimensional evaluation. For retrieval effectiveness, we employ two key metrics: $Relevance$ measures how well a retrieved memory matches a query and is typically calculated by having the LLM judge each entry's relevance, then dividing the number of relevant entries by the total retrieved. $Accuracy$ is evaluated by comparing retrieved documents to a ground truth set and metrics such as recall and F1 scores can often be used. For retrieval efficiency, we use the average retrieval time \textbf{per memory token} $T_{retr}$. For the memory formation process, we focus on energy consumption $E_{token\_m}$ and processing time $T_{cost}$ per memory token. 

Based on these metrics, we design two comprehensive evaluation indicators that quantify the energy cost required to achieve a unit improvement in retrieval relevance and accuracy:

\textbf{Retrieval Energy-to-Relevance Ratio (RERR) }
\begin{equation}
RERR = \frac{E_{token\_m}}{Relevance}
\end{equation}

\textbf{Energy-to-Accuracy Ratio (EAR) }
\begin{equation}
EAR = \frac{E_{token\_m}}{Accuracy}
\end{equation}


\subsubsection{Trade-offs between Memory Reading, Utilization and Generation}
\paragraph{a. For Correct Generations}
\label{For Correct Generations}
For correct generations of the query $Q$ (token number = $T_q$), we propose an ideal case scenario that achieves accurate answers through minimal necessary data (as shown in Figure ~\ref{fig:memory module}) and operations: in this optimal case, the system would use only the original query with vector embeddings, employing sparse or dense retrieval to extract precisely all relevant ground truth memories $M_G$ (token number = $T_{mg}$) and no irrelevant memories, which are then combined with the original query to directly generate the ground truth answer $A_G $(token number = $T_{ag}$) through the LLM, without requiring any additional enhancement operations (such as query optimization, reranking or compression).

In the optimal case, the essential energy consumption comprises two components: retrieval cost ($E_{retr}$) and response generation cost $E_g= e_K(T_q+T_{mg}, T_{ag})$. Thus, the optimal total energy consumption can be expressed as: 
\begin{equation}
E_{optimal} = E_{retr} + E_g
\end{equation}
However, in real cases, to ensure generation quality, various additional memory reading and utilization operations in ~\ref{sec:Memory Reading} and ~\ref{sec:Memory Utilization} are often necessary. We define $E_{real}$ as the cumulative energy consumption across all stages of the generation process, for example, when using all operations,
$E_{real}=E_{retr}+E_1+E_2+E_3+E_4+E_5$.

To quantify the energy efficiency of the generation process, we propose: 

\textbf{Generation Energy Optimality Ratio (GEOR)}
\begin{equation}
GEOR = \frac{E_{optimal}}{E_{real}}
\end{equation}

This ratio serves as an indicator of how close the system's actual energy consumption is to the theoretical minimum. A GEOR value closer to 100\% indicates higher energy efficiency, suggesting that the system's energy consumption approaches the ideal case, while lower values indicate potential areas for energy optimization in the generation pipeline.

\paragraph{b. For All Generations}
The effectiveness and latency of response generation are significantly influenced by memory reading and utilization methods. For generation effectiveness, we employ three key metrics: $Relevance$ measures how well the generated response aligns with the intent and content of the initial query $Q$, typically calculated as a normalized score assigned by an LLM judge based on predefined assessment criteria. $Faithfulness$ evaluates the consistency between the generated response and the retrieved memories $M_q$, calculated as the ratio of main points in the answer attributable to the context to the total main points in the answer.
$Correctness$ assesses the accuracy of the generated response against ground truth $A_G$, measured using standard NLP metrics such as ROUGE, BLEU, or BERTScore.
For generation efficiency, we measure the average $latency$ per response token and energy consumption per token $E_{token\_g}$ which is the division of $E_{real}$ by \textbf{response token} number. Similarly, there are three energy efficiency ratios:

\textbf{Generation Energy-to-Relevance Ratio (GERR)}
\begin{equation}
GRER = \frac{E_{token\_g}}{Relevance}
\end{equation}
 
\textbf{Energy-to-Faithfulness Ratio (EFR)}
\begin{equation}
FER = \frac{E_{token\_g}}{Faithfulness}
\end{equation}

\textbf{Energy-to-Correctness Ratio (ECR)}
\begin{equation}
CER = \frac{E_{token\_g}}{Correctness}
\end{equation}

These ratios quantify the energy cost required to achieve unit improvement in each generation effectiveness dimension. Lower values indicate better energy efficiency in achieving the respective effectiveness goals.

\section{Experiments}
\subsection{Experiment settings}
\textbf{Datasets.} For Internal Memory, we treat the task as that of a daily assistant agent and utilize the LoCoMo \cite{maharana2024evaluating} dataset, which comprises very long-term dialogues. For External Memory, we use HotpotQA \cite{yang2018hotpotqa} and MuSiQue \cite{trivedi2022musique}, which are well-known datasets for evaluating multi-hop and complex reasoning abilities in retrieval-augmented tasks. Additionally, we incorporate the AI ArXiv dataset \cite{calam_ai_arxiv}, with 100 pairs of queries and answers on topics of RAG extracted by Eibich et al. \cite{eibich2024aragog}. We follow the same node splitting configuration as in Eibich's experiments. For all datasets we sampled some data and the characteristics of these sub-datasets are shown in Table ~\ref{tab:Dataset Characteristics}.
\begin{table}[htbp]
\caption{Comparison of Dataset Characteristics}
\centering
\begin{tabular}{l cc ccc c l}
\toprule
\multirow{2}{*}{Dataset} & \multicolumn{2}{c}{Memory Node} & \multicolumn{3}{c}{Query} & Answer & \multirow{2}{*}{Scenario} \\
\cmidrule(lr){2-3} \cmidrule(lr){4-6} \cmidrule(lr){7-7}
& Number & Size* & Types & Difficulty & Length & Length & \\
\midrule
LoCoMo & 419 & 54.6 & 5 & Easy & Short & Short & Daily Life \\
\cdashline{1-8}[1pt/1.5pt]
\addlinespace
HotpotQA & 992 & 128.5 & 1 & Hard & Short & Short & Open-Domain \\
\addlinespace
MuSiQue & 2000 & 111.7 & 1 & Hard & Short & Short & Open-Domain \\
\cdashline{1-8}[1pt/1.5pt]
\addlinespace
AI ArXiv & 51270 & 27.5 & 1 & Medium & Long & Long & Academic \\
\bottomrule
\multicolumn{8}{l}{\small *The size of memory node is the average token number.}\\
\label{tab:Dataset Characteristics}
\end{tabular}
\end{table}
\vspace{-10pt}

\textbf{Metrics tools.}
We evaluate Retrieval-Relevance, Generation-Relevance and Faithfulness using UpTrain’s \cite{UpTrain} LLM-based scorer with the GPT-4 API. For Retrieval-Accuracy we measure the Mean Reciprocal Rank (MRR) and Recall based on the ground truth provided by the datasets, while Generation-Correctness is measured using Recall-Oriented Understudy for Gisting Evaluation (ROUGE) \cite{lin2004rouge}. All score ranges from 0 to 100.

\textbf{Hardware.}
To evaluate LLM performance across different resource scenarios, we configured two hardware setups. The \textbf{resource-constrained environment} utilizes a single NVIDIA GeForce RTX 4080 GPU(16GB, 320W TDP) and an Intel i7-13700K processor (16-core 24 threads, 125W TDP). For the \textbf{resource-abundant environment}, we employed a cluster of eight NVIDIA A100 GPUs (80GB, 400W TDP) and two AMD EPYC 7J13 processors (64-Core 128 threads, 280W TDP), with NVIDIA NIM inference accelerating microservices \footnote{\url{https://build.nvidia.com/models}}, representing advanced high-performance computing capabilities compared with resource-constrained setup.
Tasks 1 and 2 are conducted on the resource-constrained setup, while Task 3 leverages the resource-abundant configuration.

\textbf{Energy measurement tools.}
For hardware energy measurement, we employed two widely-used tools: powercap and pynvml. Powercap \cite{powercap2024} interfaces with RAPL (Running Average Power Limit), which is supported by both our Intel and AMD processors, to collect CPU energy consumption data through records in system files. For GPU energy monitoring, we utilized pynvml \cite{pynvml2024}, the Python binding for NVIDIA Management Library (NVML), which provides direct access to GPU power sensors. Both tools offer millisecond-level sampling precision with minimal overhead to the system being measured.

\textbf{LLM selection.}
In the {resource-constrained environment}, we select the Llama-3.1-8B-Instruct model with 8-bit quantization.
In the resource-abundant environment, we deploy the Llama-3.1-8B-Instruct on one A100 and Llama-3.1-70B-Instruct on four A100.

\subsection{Task 1: Memory Formation and Retrieval}
In this experiment, we selected 22,865 tokens of ground-truth documents from LoCoMo (Internal Mem), three different sized subsets (22,226, 99,946 and 202,676 tokens) from MuSiQue (External Mem). We use LlamaIndex \cite{LiuLlamaIndex2022} for different indexing implementations and the retrieval top-K is set to 5. Ten experiments were repeated to take the average value.
\paragraph{a. Different dataset, same number of tokens}
Table ~\ref{Memory Formation Operations (different dataset, same number of tokens)} shows the energy and time cost comparison of different indexing methods for different datasets with similar number of total memory token counts.

\textit{Energy Cost Per Token.}
The pattern of energy cost per token reveals two critical influencing factors. The indexing method shows substantial impact, with Vector-based (\textbf{non-LLM}) indexing demonstrating significantly lower energy consumption (\boldmath{$10^{-2}$ J/token}) compared to \textbf{LLM-dependent methods} (Keyword, Summary, and Graph), which require approximately \boldmath{$10^{+1}$ J/token}. Moreover, despite similar total token counts, the memory configuration significantly affects energy efficiency, with the internal memory setup (419 smaller nodes) consistently consuming more energy than the external configuration (220 larger nodes) across all indexing methods.

\textit{RERR and EAR.}
The RERR and EAR cost metrics derive from the energy per token and essentially reflect the large differences between the different indexing approaches, with vector indexing achieving \textbf{greater per-unit retrieval effectiveness} at significantly lower energy cost 
 (\boldmath{$10^{-4}$J}) compared to LLM-dependent approaches (\boldmath{$10^{-1}$J}). 
A distinctive pattern emerges in the cost distribution: internal data exhibits higher RERR and lower EAR values, while external memory shows the opposite trend, indicating varying energy investments required for achieving optimal Relevance and Accuracy across different scenarios. Notably, the \textbf{Graph} index, despite being the most sophisticated method, demands higher energy costs in simpler query scenarios (internal, 0.717 J and 1.035 J) compared to complex multi-hop queries (external, 0.565 J and 0.390 J), emphasizing the importance of context-appropriate index selection rather than defaulting to the most advanced option to avoid unnecessary energy expenditure.

\textit{Time Cost.}
The time cost reveals that node configuration significantly impacts both writing and reading performance, with the internal memory's more nodes requiring longer processing times for both indexing and querying compared to external data. The Vector-based method, operating without LLM, demonstrates extremely efficient processing times (\boldmath{$10^{-5}$ s/token}) for the moderate token count (around 22k), while LLM-dependent methods demand substantially longer indexing times, highlighting the need to consider these significant preprocessing costs in specific use scenarios. 
Additionally, certain indexing methods show reading efficiency advantages in specific contexts, such as \textbf{Keyword}'s faster reading times (2.01e-06 s/token), while advanced methods like \textbf{Graph} indexing require much longer (4.55e-02 s/token), suggesting that optimal time efficiency requires careful matching between indexing strategies and application requirements rather than assuming more sophisticated methods will always yield better performance.

\begin{table}[htbp]
    \centering
    \small
    \setlength{\tabcolsep}{4pt}
    \caption{Cost Comparison of Memory Formation (different dataset, similar number of tokens)}
    \begin{tabular}{l*{11}{c}}
    \toprule
    \multirow{3}{*}{Index} & \multicolumn{5}{c}{Internal Mem, Token Num: 22,865, Node num: \textbf{419}} &  \multicolumn{5}{c}{External Mem, Token Num: \textbf{22,226}, Node Num: \textbf{220}} &   \\
    \cmidrule(lr){2-6} \cmidrule(lr){7-11}
    & RERR & EAR & \makecell{Writing\\Time (s)} & \makecell{Reading\\Time (s)} & \makecell{Energy per\\token (J)}  
    & RERR & EAR & \makecell{Writing\\Time (s)} & \makecell{Reading\\Time (s)} & \makecell{Energy per\\token (J)} & \\
    \midrule
    Vector & \textbf{1.79e-04} & \textbf{2.12e-04} & \textbf{8.75e-05} & \textbf{3.29e-05} & 1.18e\textbf{-02} &\textbf{2.25e-04} &\textbf{1.01e-04} &\textbf{5.14e-05} &1.59e-05 & 6.12e\textbf{-03} \\
    \cdashline{1-11}[1pt/1.5pt]
    \addlinespace
    Keyword & \textbf{8.15e-01} & 10.23e-01 & 3.02e-01 & 8.88e-05 & 3.27e\textbf{+01} &\textbf{9.60e-01} &5.15e-01 &1.75e-01 &\textbf{2.01e-06} &2.52e\textbf{+01} \\
    \addlinespace
    Summary & 6.21e-01 & \textbf{12.67e-01} & 3.10e-01 & 4.14e-04 & 3.40e\textbf{+01} &9.19e-01 &\textbf{5.24e-01} &1.71e-01 &8.77e-05 & 2.50e\textbf{+01}\\
    \addlinespace
    Graph & \underline{7.17e-01} & \underline{10.35e-01} & 3.29e-01 & \textbf{4.86e-02} & 3.63e\textbf{+01} &\underline{5.65e-01}  &\underline{3.90e-01} &1.77e-01 &\textbf{4.55e-02} & 2.57e\textbf{+01} \\
    \bottomrule
    \end{tabular}
    \label{Memory Formation Operations (different dataset, same number of tokens)}
\end{table}

\paragraph{b. Same dataset, different number of tokens}
Table ~\ref{Memory Formation Operations (same dataset, different number of tokens)} and External Mem part in Table ~\ref{Memory Formation Operations (different dataset, same number of tokens)}  shows the cost comparison of different indexing methods for same datasets with increasing number of memory tokens. The cost pattern analysis across increasing token numbers (from 22,226 to 99,946 to 202,676) in external memory reveals a trend in efficiency scaling. The Vector-based method shows \textbf{consistently increasing costs} across all metrics: energy per token rises from 6.12e-03 J to 9.80e-03 J, while its RERR and EAR values also increase proportionally. 
In contrast, LLM-dependent methods demonstrate relatively \textbf{stable cost patterns}, remaining nearly constant metrics across different token counts.
The narrowing gap and the stability of the LLM-dependent approach's performance with increasing token volumes suggests its potential to remain cost-effective at scale. Nonetheless, the huge magnitude difference (\boldmath{$10^{-3}$} vs \boldmath{$10^{+1}$}) in energy costs between the two still needs to be addressed first.

\begin{table}[htbp]
    \centering
    \small
    \setlength{\tabcolsep}{4pt}
    \caption{Cost Comparison of Memory Formation (same dataset, different number of tokens)}
    \begin{tabular}{l*{11}{c}}
    \toprule
    \multirow{3}{*}{Index} & \multicolumn{5}{c}{External Mem, Token Number = \textbf{99,946}} & \multicolumn{5}{c}{External Mem, Token Num = \textbf{202,676}} &  \\
    \cmidrule(lr){2-6} \cmidrule(lr){7-11} 
    & RERR & EAR & \makecell{Writing\\Time (s)} & \makecell{Reading\\Time (s)} & \makecell{Energy per\\token (J)}  
    & RERR & EAR & \makecell{Writing\\Time (s)} & \makecell{Reading\\Time (s)} & \makecell{Energy per\\token (J)} & \\
    \midrule
    Vector & 2.48e-04 & 1.12e-04 & 5.16e-05 & 4.38e-05 &\textbf{8.94}e-03 &2.53e-04 & 1.28e-04 & 5.40e-05 & 7.03e-05 & \textbf{9.80}e-03 &  \\
    \cdashline{1-11}[1pt/1.5pt]
    \addlinespace
    Keyword & 9.79e-01 & 6.40e-01 & 1.79e-01 & 4.76e-06& \textbf{2.50}e+01 & 9.23e-01& 6.59e-01& 1.71e-01& 9.56e-06& \textbf{2.48}e+01 &  \\
    \addlinespace
    Summary & 7.87e-01 & 4.94e-01 & 1.76e-01 & 2.12e-04&\textbf{2.52}e+01 &7.66e-01 & 4.81e-01 &1.69e-01 & 3.73e-04 & \textbf{ 2.50}e+01& \\
    \addlinespace
    Graph & 6.11e-01 & 3.21e-01 & 1.79e-01 & 4.29e-02&\textbf{2.57}e+01 &6.17e-01 &3.23e-01 &1.77e-01 &3.58e-02 & \textbf{2.59}e+01& \\
    \bottomrule
    \end{tabular}
        \label{Memory Formation Operations (same dataset, different number of tokens)}
\end{table}

\subsection{Task 2: Memory Reading, Utilization and Generation} \label{sec:Task 2}
\paragraph{a. For all generations}

In this experiment, we selected 100 simpler, non-multi-hop reasoning queries from the LoCoMo, 200 queries from HotpotQA \& MuSiQue, and 100 academic queries from ArXiv dataset. The indexing and retrieval are fixed to the vector-based method and the retrieval top-K is set to 5. Ten experiments were repeated to take the average value. The results are shown in Table~\ref{Comparison of Memory Reading, Utilization Operations} and Table~\ref{tab:comprehensive-cost-ratio}, where Table 5 is a multiples analysis based on the GERR, EFR, and ECR values in Table 4.
For the choice of \textit{query optimisation} methods, we followed the experimental results of other previous RAG evaluation work \cite{wang2024searching}, choosing the HyDE method which has relatively better metrics. For the other operations, we use prompt engineering for ordinary LLM inference to implement them.
\begin{table}[htbp]
    \centering
    \small
    \setlength{\tabcolsep}{3pt}
    \caption{Cost Comparison of Memory Reading, Utilization Operations}
    \begin{tabular}{l*{12}{c}}
    \toprule
    \multirow{2}{*}{\textbf{Operation}} & \multicolumn{4}{c}{\textbf{Internal Mem.}} & \multicolumn{4}{c}{\textbf{External Mem. A}} & \multicolumn{4}{c}{\textbf{External Mem. B}} \\
    \cmidrule(lr){2-5} \cmidrule(lr){6-9} \cmidrule(lr){10-13}
    & GERR & EFR & ECR & latency(s) & GERR & EFR & ECR & latency(s) & GERR & EFR & ECR & latency(s) \\
    \midrule
    Retrieval + Generation & 0.144&0.096 &0.328 &0.059&\textbf{0.951} &0.102 &\textbf{1.440} &0.059 &  0.230 &\textbf{0.176} &0.669 &\textbf{0.102} \\
    \cdashline{1-13}[1pt/1.5pt]
    \addlinespace
    + retrieval detection &0.188 &0.137 &0.413 &0.077&\textbf{1.114} &0.120 &\textbf{1.686} &0.069 &0.385 & \textbf{0.294}& 1.120&\textbf{0.111} \\
    \addlinespace
    + query optimization &5.005 &\textbf{3.312} &11.918 &\textbf{2.115} &\textbf{15.957} &1.823 &\textbf{19.490} &1.117 & 2.806& 2.012& 8.135&1.073 \\
    \addlinespace
    + reranking &0.584 &\textbf{0.413} &1.307 &\textbf{0.245} &\textbf{3.172} &0.356 &\textbf{4.490} &0.202 &0.499 &0.361 & 1.368& 0.192\\
    \addlinespace
    + compression &2.542 &1.686 &5.261 &1.084 &\textbf{12.884} &\textbf{2.039} &\textbf{23.653} &\textbf{1.187} &1.193 & 0.942&3.403 &0.468 \\
    \midrule
    + all &6.561 & \textbf{4.604} &14.325 & \textbf{2.841} &\textbf{28.568} &3.596 &\textbf{43.460} &2.106 &3.433 &2.809 &9.637 &1.330 \\
    
    \bottomrule
    \end{tabular}
    \label{Comparison of Memory Reading, Utilization Operations}
\end{table}
\vspace{-10pt}

\begin{table}[htbp]
\centering
\caption{Energy Cost Multiples Analysis (\textit{Retrieval + Generation} as Baseline)}
\begin{tabular}{lccc}
\toprule
\textbf{Operation} & \textbf{Internal Mem.} & \textbf{External Mem. A} & \textbf{External Mem. B} \\
\midrule
+ retrieval detection & 1.30$\times$ & 1.17$\times$ & 1.67$\times$ \\
+ query optimization & 35.19$\times$ & 15.95$\times$ & 11.93$\times$ \\
+ reranking & 4.11$\times$ & 3.31$\times$ & 2.09$\times$ \\
+ compression & 17.07$\times$ & 16.45$\times$ & 5.21$\times$ \\
\midrule
+ all & \textbf{45.70}$\times$ & \textbf{31.74}$\times$ & \textbf{15.08}$\times$ \\
\bottomrule
\multicolumn{4}{l}{\small Note: Values represent geometric mean multiples of relative increases in GERR, EFR, and ECR}\\
\end{tabular}
\label{tab:comprehensive-cost-ratio}
\end{table}


\textit{External Mem B (ArXiv).} The arXiv dataset, characterized by longer queries and a large number of nodes (51,270), impacted retrieval and detection costs, leading to \textbf{relatively high latency} (0.102, 0.111). Longer responses contained more information and divergence points, while smaller node sizes reduced the information in individual text fragments. This mismatch made it harder for the LLM to align responses with retrieved text, leading to \textbf{higher EFR cost} (0.176, 0.294).

\textit{External Mem A (HotpotQA \& MuSiQue).} 
The multi-hop query introduced higher query complexity, which notably led to \textbf{high GERR and ECR metrics}. These results highlighted the additional energy costs required to improve response-query relevance and ensure response accuracy. Although the \textit{compression} effectively alleviated the pressure of larger nodes on LLM input length, it incurred a significant energy cost (a \textbf{16-fold increase}), prompting a need to carefully weigh the cost-effectiveness of such an expensive optimization step in various scenarios.

\textit{Internal Mem (LoCoMo).}
LoCoMo consisted of simple, everyday conversational queries. Its low query complexity was reflected in the relatively \textbf{low initial GERR} (0.144) and \textbf{ECR} (0.328). However, during \textit{query optimization}, the generated hypothetical documents for these simple tasks often introduced redundant or distracting information—particularly in time-, location-, or person-based dialogues. This "over-optimization" led to a dramatic increase in energy consumption and latency (about \textbf{36-fold rise}), underscoring the critical importance of carefully selecting operational strategies.

\begin{figure}[htbp]
    \centering
    \begin{subfigure}[b]{0.48\linewidth}
    \centering
    \includegraphics[width=\linewidth]{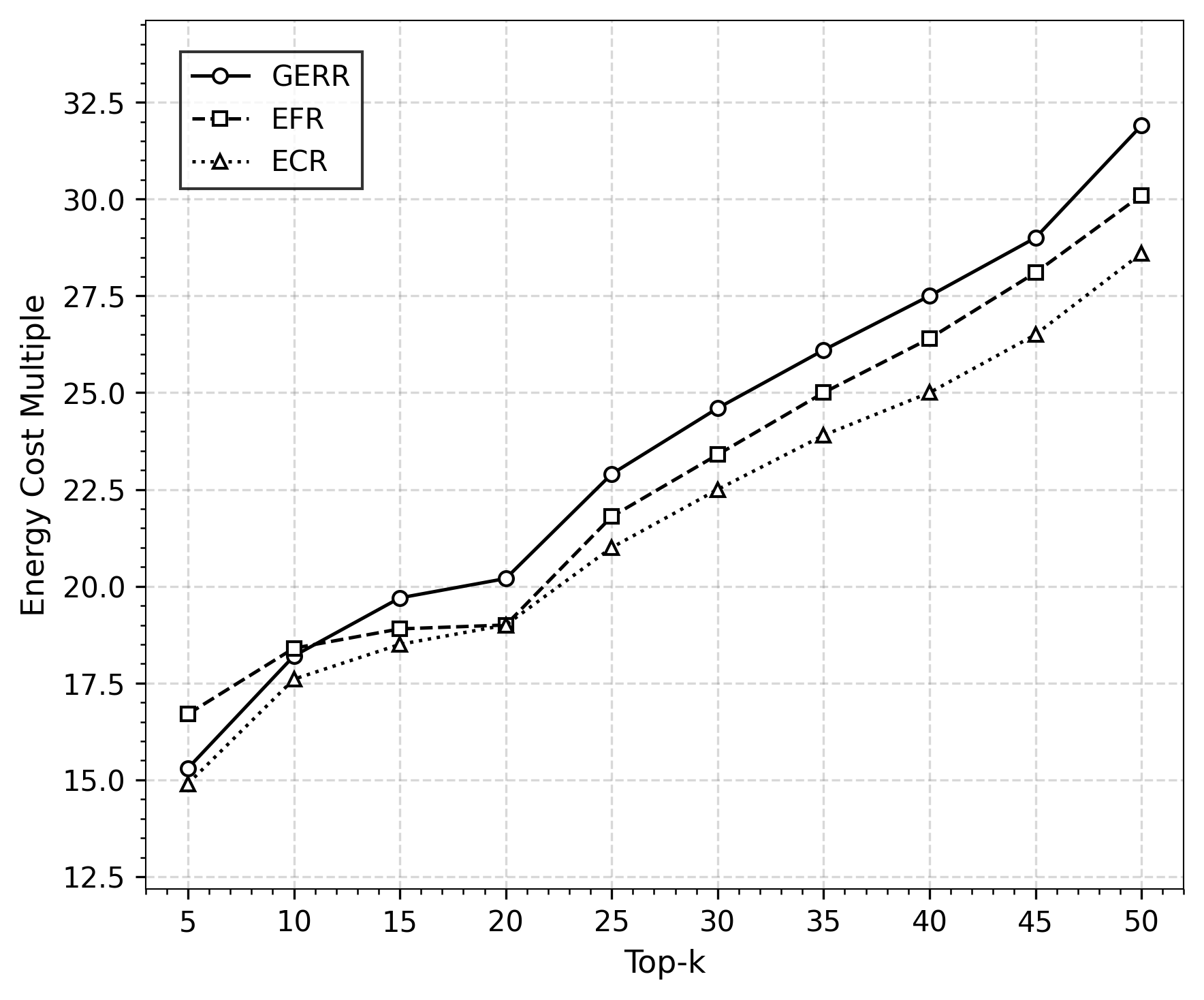}
    \caption{Cost Multiple Comparison (External Mem. B)}
    \label{fig:cost_ratio_comparison}
\end{subfigure}
    \hfill 
    \begin{subfigure}[b]{0.48\linewidth}
        \centering
        \includegraphics[width=\linewidth]{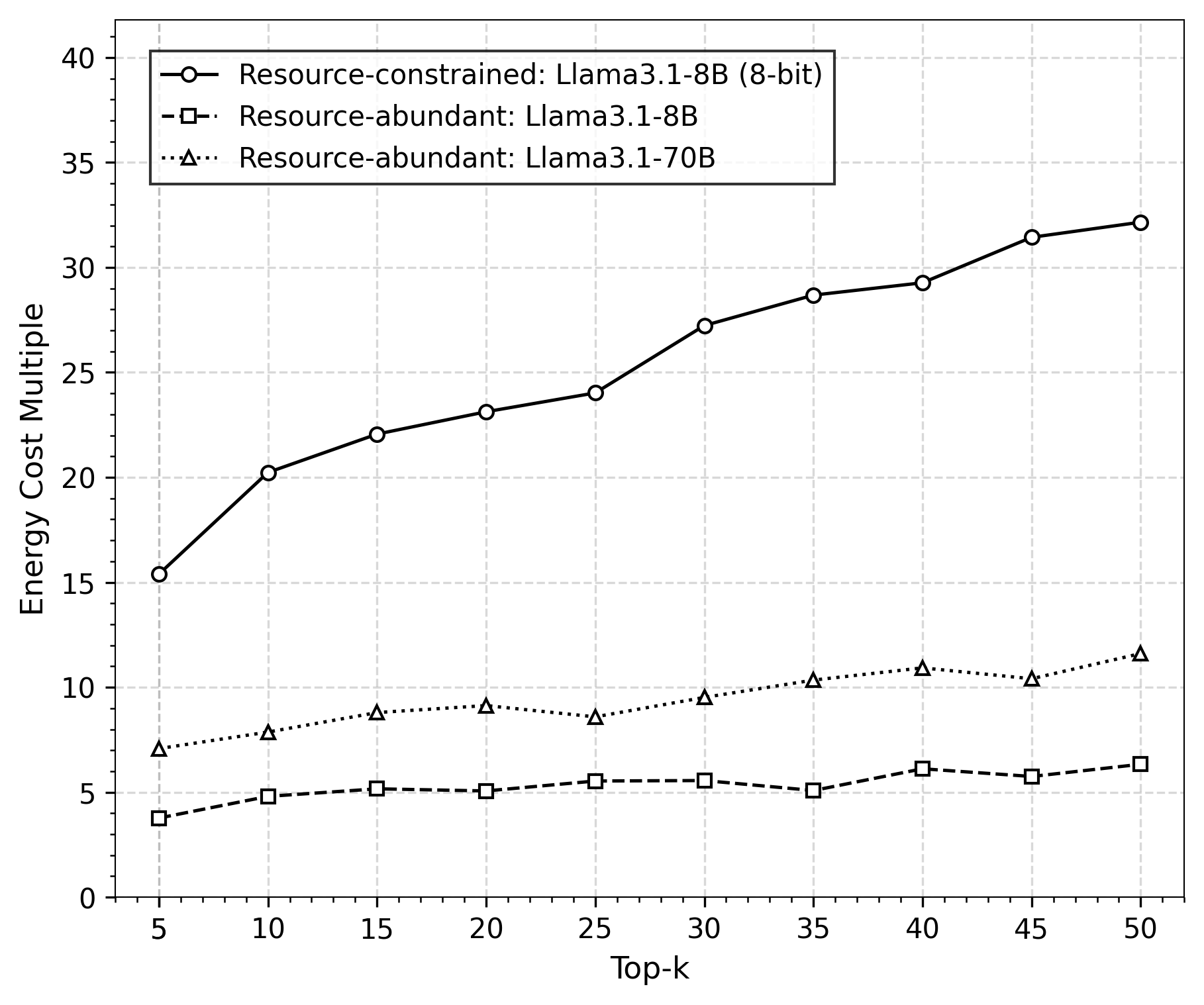}
        \caption{Cost Multiple Comparison (Different Models)}
        \label{fig:relative_performance}
    \end{subfigure}
    \caption{The Relationship between Energy Cost Multiple and Memory Reading Workload (top-K value)}
    \label{fig:comparison}
\end{figure}

\textit{Costs of different operations.}
With relatively low additional costs (1.17 to 1.67-fold increase), \textit{retrieval detection} can be acceptable as a better trade-off operation, if in scenarios with simpler queries or where the LLM’s own knowledge is sufficient, timely detection can prevent unnecessary downstream operations. In contrast, sub-optimal operations can have extreme energy impacts, an example of which is \textit{query optimisation} in internal memory. All operations (+all) resulted in energy costs increasing by \textbf{an order of magnitude or more} (45.7-fold, 31.74-fold, 15.08-fold). Although the relative increase of External Mem.B was smaller, the overhead \textbf{linearly scaled} as the memory reading workload (top-K retrieved documents) increased, as shown in Figure ~\ref{fig:cost_ratio_comparison}.  
\vspace{-15pt}

\paragraph{b. For correct generations}
In this experiment, we selected 100 correctly answered queries each from the LoCoMo, HotpotQA and MuSiQue respectively. Through 10 experimental iterations with 100 different (input length, output length) pairs in our \textbf{resource-constrained environment}, we measured the actual energy consumption using monitoring tools and then applying OLS regression to fit the energy model ~\ref{sec: problem}. We derived the coefficients {\boldmath$\alpha_{K,0} = 0.042933$}, {\boldmath$\alpha_{K,1} = 9.109322$}, and {\boldmath$\alpha_{K,2} = 0.000513$} for the equation (3), where K represents Llama-3.1-8B-Instruct model with 8-bit quantization. The high R-squared value of 0.989 validates that this model effectively captures the energy-token numbers relationship.

Based on our defined minimal essential operations and data in ~\ref{For Correct Generations}, we calculated $E_g$ by substituting into equation (3) and measured the retrieval cost $E_{retr}$ and real energy cost $E_{real}$ when all operations were performed. Following Equations (6) and (7), we calculated the Generation Energy Optimality Ratio (GEOR) for the three datasets. The results show that the GEOR values are extremely low as shown in Table ~\ref{GEOR}, all only \textbf{about 1-2\%}. Several factors lead to excessive extra energy consumption, yielding these exceptionally low values: the retrieval and processing of many irrelevant memories, the accumulated overhead from additional operations, and the extended processing time from retrieval and computation latency. This indicate that while the LLM agent is capable of correctly answering memory-based tasks, there is a considerable gap between the system’s current energy performance and the theoretical minimum, underscoring the inefficiency in energy utilization during the generation pipeline, as well as the significant room for improvement in optimising the energy efficiency of the LLM agent's design pattern.

\vspace{-10pt}
\begin{table}[htbp]
\centering
\caption{Generation Energy Optimality Ratio (when all operations are performed)}
\begin{tabular}{lccc}
\toprule
\textbf{Dataset} & \textbf{LoCoMo} & \textbf{HotpotQA} & \textbf{MuSiQue} \\
\midrule
GEOR & 1.378\% & 1.012\% & 1.456\% \\
\bottomrule
\end{tabular}
\label{GEOR}
\end{table}

\subsection{Task 3: Resource-constrained vs. Resource-abundant Environment}
\begin{figure}
    \centering
    \includegraphics[width=1\linewidth]{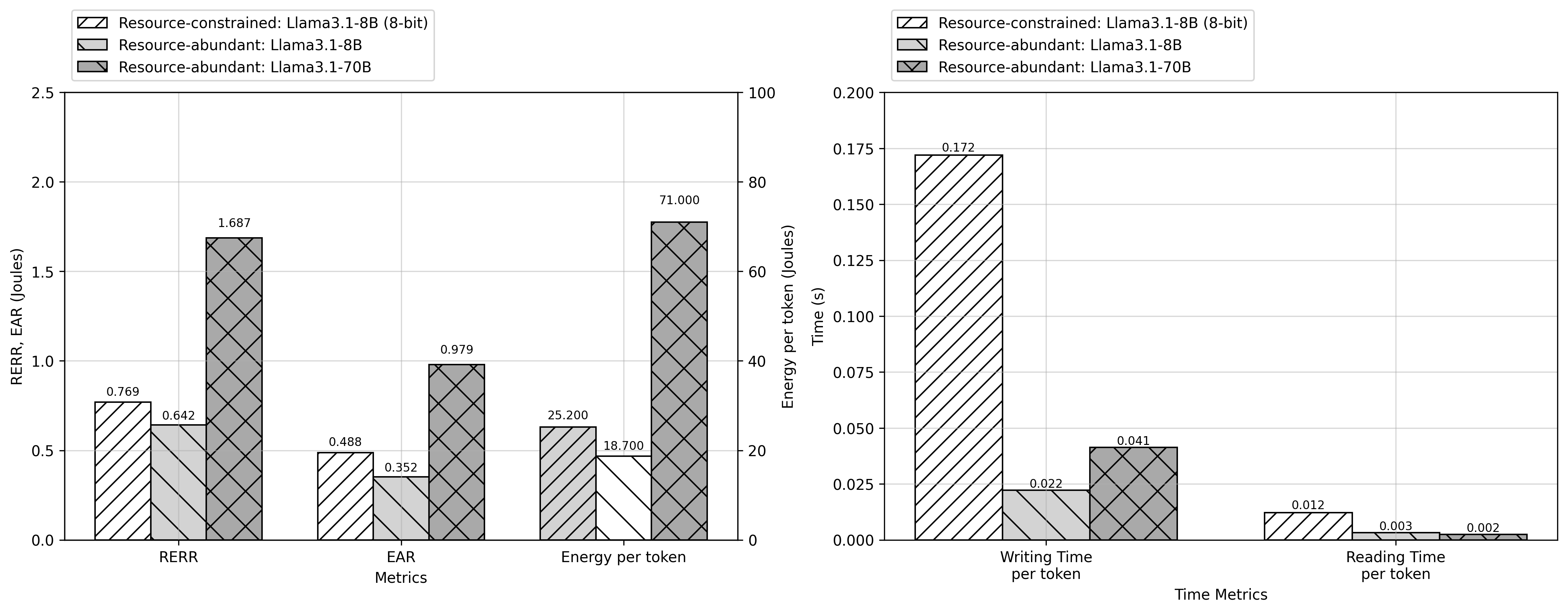}
    \caption{Comparison of Memory Formation in Resource-constrained vs. Resource-abundant Environments}
    \label{fig:multi-model}
\end{figure}

In this part we repeated the same experiments for Task 1 and 2 in resource-abundant environment using Llama-3.1-8B-Instruct and Llama-3.1-70B-Instruct. The results shown in Figure ~\ref{fig:multi-model} represent the average metrics of the three LLM-based index methods (Keyword, Summary, and Graph) for the Memory Formation and Retrieval phases, and Figure ~\ref{fig:relative_performance}, the energy of the Generation phases for performing all operations on the Arxiv dataset compared to the multiples of energy consumption of the baseline (Retrieval + Generation).

\textit{Cost-Effectiveness Paradox.} Despite operating with reduced precision (8-bit quantization), the resource-constrained environment shows notably poor cost-efficiency ratios. The resource-abundant environment, running full-precision 8B model, achieves lower RERR (0.642 vs 0.769) and EAR (0.352 vs 0.488), and consumed less energy per memory token (18.7 vs 25.2 joules). Given that RERR and EAR measure the energy cost required to achieve unit retrieval relevance and accuracy respectively, this counter-intuitive result reveals that resource-constrained environments are forced to pay higher operational costs (35\% more energy per token) while achieving lower effectiveness.
    
\textit{Model Size Scaling Benefits.} The 70B model consumes approximately three times more energy per memory token than the quantized 8B model (67.5 vs 25.2 joules), while its RERR (1.687 vs 0.769) and EAR (0.979 vs 0.488) are only about twice as high. This disproportionate relationship indicates that the 70B model achieves higher relevance and accuracy improvements relative to its energy consumption increase. This advantageous scaling property - where performance improvements outpace energy cost increases - is exclusively available to resource-abundant environments.

\textit{Service Quality Disparity.} The most striking disparity appears in processing efficiency. The resource-constrained environment exhibits dramatically slower indexing speeds, with writing time per token (0.172s) being approximately 8 times slower than the resource-abundant environment (0.022s for 8B model), while the difference in read time is also four to six times greater. This difference in processing speed signals that resource-constrained environments are paying for longer preprocessing times but gaining a more sub-optimal quality of service, potentially rendering these LLM agent systems impractical for real-world applications in resource-limited settings.

\textit{Optimization Penalty Paradox.} Further analysis of the Generation phase (Figure ~\ref{fig:relative_performance}) reveals an even more severe energy efficiency gap with optimization operations (reranking, compression, etc.). While all configurations show increasing energy costs as the workload grows, the resource-constrained environment exhibits a dramatically steeper curve - from 15x to 32x baseline consumption as top-K increases from 5 to 50. In comparison, resource-abundant environments show more moderate increases (from 4x to 6x for 8B, and 7x to 12x for 70B). This divergent scaling behavior indicates that although the baseline energy overhead might be lower in resource-constrained environments, their efficiency deteriorates much more rapidly with increased context length and additional processing steps. This creates a paradoxical situation where attempts to improve system performance through optimization operations actually widen the energy efficiency gap between resource-abundant and resource-constrained deployments.

\section{Discussions and Conclusions}
Our experimental investigation into LLM agent systems provides empirical evidence for the Sustainable AI Trilemma and its constituent dilemmas, revealing how the tensions between capability, digital equity, and environmental sustainability manifest in practical deployments. Through a systematic examination of memory operations in LLM agents, we observe several critical patterns that validate and deepen our understanding of these challenges.

The \textbf{Operational Dilemma} manifests through stark energy consumption patterns in LLM agent systems. The fundamental tension between \textbf{scalability and sustainability} is evident in the orders-of-magnitude difference between LLM-dependent and non-LLM methods, with optimization attempts often leading to significant energy penalties. The remarkably low Generation Energy Optimality Ratio (GEOR) of 1-2\% highlights how current architectures operate far from optimal efficiency, suggesting that without fundamental breakthroughs in energy efficiency, the widespread deployment of LLM agents may face significant sustainability barriers.

The \textbf{Evaluation Dilemma} reflects the crucial transition from \textbf{mere awareness to precise measurement} of AI's environmental impact. While previous approaches relied on simple proxies like the product of GPU power and runtime, our evaluation framework introduces targeted metrics and more concrete evidences that reveal nuanced trade-offs and previously obscured patterns between energy costs and system performance. This progress in measurement capability, though significant, also highlights the continuing challenge of developing comprehensive assessment tools for increasingly complex AI applications.

The \textbf{Design Dilemma} challenges the prevailing "LLM-centric" paradigm that delegates complete operational control to LLM. Our findings reveal the inefficiencies and overhead associated with \textbf{fully LLM autonomous} operation while only some LLM operations can be \textbf{cost-effective} like retrieval detection. The systematic increase in energy costs with task complexity suggests the need for a more balanced approach that strategically combines LLM and traditional low-cost non-LLM methods. In addition, whether LLMs can accurately assess the boundaries of their capabilities \cite{ren2023investigating}, adaptively avoid inappropriate operations and perform optimal operation allocation are all key issues that need to be further explored.

The \textbf{Access Dilemma} emerges as a profound manifestation of digital inequality, where the relationship between resources \textbf{barriers and benefits} creates a negative feedback loop of technological disparity: limited computational resources lead to both higher operational costs and reduced capabilities but also experience disproportionate penalties when attempting to optimize their systems. This troubling "technological poverty trap" creates a fundamental barrier to AI democratization, where the attempts to improve accessibility through techniques like model quantization and RAG optimization may widen the capability gap between resource-abundant and resource-constrained deployments, effectively preventing disadvantaged communities from realizing the transformative potential of AI technologies.  This pattern suggests that the current trajectory of AI development may inadvertently amplify existing digital divide rather than bridge them.

This work has highlighted the complex, interconnected challenges in developing sustainable AI systems, particularly for LLM agents. Our findings demonstrate that addressing any single aspect of sustainability often creates ripple effects across technical, social, and environmental dimensions, necessitating a holistic approach to solution design. Looking ahead, our research will focus on developing practical frameworks for sustainable AI practices that can effectively reduce environmental impact while preserving the utility of LLM agents. Through continued investigation of responsible AI design principles and their implementation, we aim to enable the equitable deployment of these systems across diverse contexts, including resource-constrained environments. This future direction emphasizes the critical balance between advancing AI capabilities and ensuring their long-term sustainability through coordinated technical innovation and thoughtful policy development.

\section*{Acknowledgments}  
We would like to thank the NVIDIA Academic Grant Program Award for providing computing resources. This work has also been partially funded by the ARIA Safeguarded AI TA3 programme.

\printbibliography
\end{document}